# Energy optimization and Performance Analysis of Cluster Based Routing Protocols Extended from LEACH for WSNs

M. Aslam, M. B. Rasheed, T. Shah, A. Rahim, Z. A. Khan*, U. Qasim[#], M. W. Qasim,

A. Hassan, A. Khan[$], N. Javaid

COMSATS Institute of Information Technology, Islamabad, Pakistan.
*Internetworking Program, Faculty of Engineering, Dalhousie University, Halifax, Canada.
[#]University of Alberta, Alberta, Canada.
[$]Sarhad University of Science and Information Technology, Peshawar, Pakistan.

**ABSTRACT**

An energy efficient routing protocol is the major attentiveness for researcher in field of Wireless Sensor Networks (WSNs). In this paper, we present some energy efficient hierarchal routing protocols, prosper from conventional Low Energy Adaptive Clustering Hierarchy (LEACH) routing protocol. Fundamental objective of our consideration is to analyze, how these ex- tended routing protocols work in order to optimize lifetime of network nodes and how quality of routing protocols is improved for WSNs. Furthermore, this paper also emphasizes on some issues experienced by LEACH and also explains how these issues are tackled by other enhanced routing protocols from classi- cal LEACH. We analytically compare the features and performance issues of each hierarchal routing protocol. We also simulate selected clustering routing protocols for our study in order to elaborate the enhancement achieved by ameliorate routing protocols.

## I. INTRODUCTION

Supply Contemporary progression in micro-electronics technology empowered designer to develop low cost, low power and small size sensors [1-2]. Hundreds and thou- sands of these sensors are deployed in WSNs according to the requirements of different applications. Sensor nodes are able to monitor, compute and transmit sensed information to core network. These sensors can communicate to each other and also to some external Base Station (BS) [4]. WSNs are used for both military and civil applications [3]. A wide-range of applications have been supported by WSNs, some of these accustomed applications are environmental monitoring, industrial sensing, infrastructure protection, battlefield, and temperature sensing.

Routing is one of the main challenge faced by WSNs. Complexity of routing protocols in WSNs is due to dynamic nature of nodes, computational overhead, no conventional addressing scheme, self-organization and limited transmission range of sensor nodes [2-4]. Sensor nodes have limited battery lifetime. Usually their battery cannot be replaced and recharged due to area of their deployment, so, the network lifetime depends upon the initial battery capacity of sensor nodes. A careful management of resources is needed to increase lifespan of WSNs. Quality of routing protocols depends upon the amount of data (Actual Data Signal) successfully received by BS from sensor nodes deployed in the network area. Number of routing protocols has been proposed for WSNs. These protocols are classified into three categories.

1. Flat routing protocols
2. Hierarchical routing protocols
3. Location based routing protocols

Hierarchical routing protocols are providing maximum energy efficient routing mechanisms, as discussed in [1-4], [7-13], [18-21]. Low Energy Adoptive Clustering Hierarchy (LEACH) routing protocol is acknowledged as a basic energy efficient hierarchical routing protocol. Many protocols have been derived from LEACH with help of some enhancements and applying advance routing techniques. This paper discuses and compares few hierarchical routing protocols like LEACH, LEACH-Centralized (LEACH-C), solar-aware LEACH (sleach), Multi-Hop LEACH, Mobile LEACH (M-LEACH) and LEACH-Selective Cluster (LEACH-SC). In [22], authors present a technique to select an energy efficient and shortest route for data transmission. These are all energy efficient, well-defined routing protocols. Our study also evaluates the operation mechanism of each routing protocol with the help of detailed flow chart.

Rest of the paper is organized as follows. We discuss the LEACH, LEACH-C, sLEACH, Multi-Hop LEACH, M-LEACH and LEACH-SC with all neces- sary detail in section 2. After that, we compare the features of these selected hierarchical routing protocols in section 3. In section 4 analytical comparison is given to elaborate the energy efficiency of all routing protocols. Simulation results are discussed in section 5. The last section concludes our comprehensive research work.

## II.    HIERARCHICAL ROUTING PROTOCOLS

This work is an extension of [24]. In hierarchical routing protocols whole network nodes are divided into multiple clusters. One node in each cluster plays leading rule. Cluster-Heads (CHs) are only nodes that can directly communicate to BS in clustering routing protocols. This significantly reduces the long distance transmission overhead of normal nodes because normal nodes have to transmit to closer CHs [1-3], [5], [7], [11-15]. In [23-25], authores presented the different cluster based routing protocols for energy effiecient data transmission from source to destination node. Transmission delay is another problem in WSNs when time critical data is required like health care applications. In [26-29], authors presented and evaluate the performance of energy efficient and delay aware routing protcols to maximize the energy consumption. They investigate the different techniques to reduce the transmission delay for reactive and time critical applications. detail description of some hierarchical routing protocols is discussed in following subsections.
Medium Access Control (MAC) protocols used in WSNs are low power and accurate with great accuracy of data delivery. In [30-32], authors presented the comparision and simulation results of different Medium Access Control (MAC) protocols. A different MAC techniques were evaluated and investigated to fulfill the different requirements like transmission delay, packet loss and energy efficiency.

### II.1    LEACH

LEACH is one of the earliest hierarchical routing protocol proposed for WSNs to increase lifespan of network. Sensor nodes organize themselves into clusters in LEACH. LEACH performs self-organizing and re-clustering functions for every round [1]. In every cluster one of sensor nodes acts as CH and remain- ing sensor nodes act as member nodes of that cluster. CHs collect the data from all nodes, aggregate received data and route all meaningful compress information to BS. Because of these additional responsibilities, CH dissipates more energy and if it remains CH permanently it will die very quickly, as it happens in case of static clustering. LEACH tackles this problem by adopt- ing randomized rotation of CHs to save battery of individual node [1,2]. In this way LEACH maximizes lifetime of network and also reduces the energy dissipation by compressing date before transmitting to BS.

Operation of LEACH is based on rounds, where each round normally con- sists of two phases. These are setup phase and steady state phase. In setup phase CHs and clusters are created. All nodes are managed into multiple clus- ters. Some nodes independently elect themselves as CHs without any negotiation to other nodes. CHs elect themselves on basis suggested percentage $P$ and their previous record as a CH. All nodes which were not CHs in previous
$1/p$ rounds, generate a number from 0 to 1 and if it is less then threshold $T(n)$ then these nodes become CHs. Threshold value is set through this formula.

$$T(n) = \begin{cases} \dfrac{P}{1 - P*(r \bmod \dfrac{1}{P})} & if n \in G \\ 0 & otherwise \end{cases} \quad (1)$$

Where, G is set of nodes that have not been selected as CHs in previous 1/p rounds, P is suggested percentage of CH, r is current round. If nodes become CHs in current round, these nodes will be CHs after next 1/p rounds [1-3]. This indicates that every node will serve as a CH equally and energy dissipation will be uniform throughout the network. Elected CHs broadcast their status using CSMA/CA protocol. Non-CH nodes select their CHs by comparing Received Signal Strength Indication (RSSI) of multiple CHs, from where nodes receive advertisements messages. All CHs will create TDMA schedule for their associated members in the cluster.

Steady state phase starts when clusters have been created. In this phase nodes communicate to chs, during allocated time slots otherwise nodes keep sleeping. Due to this attribute leach minimizes energy dissipation and ex- tends battery lifetime of all individual nodes. When data from all nodes of cluster have been received to ch, it will aggregate, compress and transmit to bs. Usually steady state phase is longer than setup phase. Leach network topology is shown in fig 1.

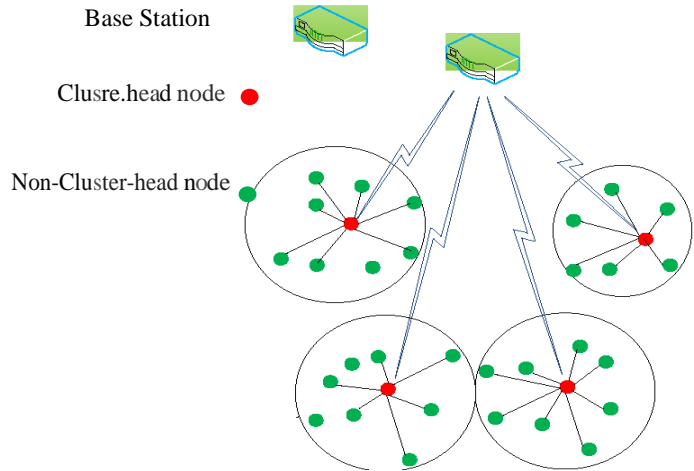

Fig. 1 Clustering topology of LEACH

## II.2 LEACH-C

Ali Although conventional LEACH has many advantages e.g, energy maximization of network and also provides limited network scalability. But LEACH does not guarantee the effective location and optimal number of CHs during all rounds [1-4], [6-7]. It is due to its distributed algorithm of clustering creation. So setup phase of LEACH needs to be modified for more effective cluster formation. For this purpose LEACH-C has been proposed by Heinzelman and co-authors in [3].

In LEACH-C during setup phase all nodes send their energy level, node IDs and location information to BS [2]. BS specifies some nodes as CHs and non- CHs with help of central control algorithm. Using central control algorithm BS compares the energy of all nodes with specific average energy level [6]. If energy of some nodes is less than average energy, BS categorizes these nodes as member nodes. BS selects optimal number of CHs from nodes having energy above than average energy level. Then BS broadcasts the node IDs of selected CHs to all networks nodes.

Central control algorithm of BS tries to minimize the distance between member nodes and CHs. In this way LEACH-C reduces the energy dissipation of member nodes because now nodes have to transmit to CH at very short distance. This central control algorithm produces much better clustering than distributed control algorithm. LEACH-C uses some necessary assumptions that each node can compute its energy, knows its location and can transmit this information to BS, no matter how much far away the BS is placed.

Because nodes can adapt multiple transmission power level that's why nodes can vary their range of communication for intra-cluster communication and inter-cluster communication [2]. Steady state phase of LEACH-C is similar to LEACH but LEACH-C enhances the number of packets received at BS. It is because of optimal number of selected CHs and their effective location with respect to non-CH nodes. LEACH-C is slightly better than LEACH, however, it has some drawbacks also like, in setup phase all nodes have to send their information to BS. This dissipates additional energy of all nodes for every round. BS selects most suitable CHs and broadcasts their node IDs to all nodes. Normal nodes also dissipate energy unnecessarily in matching their node IDs to CHs node IDs. This extra computation over-head is main disadvantage of LEACH-C.

## II.3 sLEACH

Energy harvesting is essential incase of some specific applications of WSNs, especially when sensor nodes are deployed in non-accessible areas like bat- tlefield and forest [9]. To deal with such kind of applications sLEACH has been proposed by authors in [9], in which lifetime of the WSNs has been improved through solar cell installation over nodes. In sLEACH some nodes are facilitated by solar power and these nodes will act as CHs more frequently, depending upon their solar status. Both LEACH and LEACH-C are extended by sLEACH.

### II.3.1 SOLAR-AWARE CENTRALIZED LEACH

In solar-aware Centralized LEACH, CHs are selected by BS with help of improved central control algorithm. Normally BS selects solar powered nodes as CHs because these nodes have maximum residual energy. Authors in [9], improve the conventional CH selection algorithm used in LEACH-C [2,3]. In sLEACH nodes transmit their solar status to BS along with energy level and nodes with higher energy are selected as CHs. Performance of sensor network increases when number of solar-aware nodes are increased. Sensor network lifetime also depends upon the sunDuration. It is the time when energy is har- vested. If sunDuration is smaller CH handover is performed in sLEACH [9]. If node serving as CH is running on battery and other node in same cluster sends data with a flag, denoting that its solar power is increased, this node will become CH in place of first serving CH. This new CH is selected during steady state phase, that also enhance the lifetime of the network nodes.

### II.3.2 SOLAR-AWARE DISTRIBUTED LEACH

In Solar-aware Distributed LEACH a distributed algorithm is used for cluster- ing process. In setup phase, CH's selection preference is given to solar-driven nodes. Initially probability for solar-driven nodes is higher than battery-driven nodes. Equation 1 is needed to be changed to increase the probability of solar- driven nodes. This can be achieved by multiplying a factor $sf(n)$ to right side of the equation 1.

$$T(n) = sf(n) \times p/1 - (cHeads/numNodes) \qquad (2)$$

Where $sf(n)$ is equal to 4 for solar-driven nodes, $sf(n)$ is equal to 1/4 for bat- tery driven nodes. $P$ is the desired percentage of optimal CHs. The *cHeads* is number of CHs since the start of last meta round. The *numNodes* is total number of nodes [8], [9]. Remaining setup phase portion of solar-aware Distribute LEACH is like conventional LEACH. Like solar-aware Centralized LEACH, in Steady state CH handover can be performed. If solar-power is added in non-CH node and CH is battery driven node then CH's handover is executed. Efficiency of sLEACH-Distributed can be maximized by adding more solar-driven nodes. As shown in Flow chart, setup phase is distributed and probabilistic base like LEACH but in this case probability of solar-driven node is kept higher. These solar-driven nodes can become CHs consecutively in next round also if their probability is still higher than other nodes.

### II.3.3 MULTI-HOP LEACH

When network diameter is increased beyond certain level, distance between CH and BS is increased enormously. This large network is not suitable for LEACH [11], in which BS is assumed at single-hop to all CHs. In this case transmission energy cost of CHs is not affordable. To address this problem Multi-hop LEACH is proposed in [12]. Multi-hop LEACH is another enhanced extension of LEACH to increase energy efficiency of the WSNs [11-13]. Multi- hop LEACH is also a distributed clustering routing protocol. Like LEACH, in Multi-Hop LEACH some nodes elect themselves as CHs and other nodes associate themselves with elected CHs to complete cluster formation in setup phase.

In steady state phase, CHs collect data from all their member nodes and transmit data directly or through other CHs to BS after aggregation. Multi- Hop LEACH allows two types of communication operations, inter-cluster com- munication and intra-cluster communication.

In Multi-hop inter-cluster communication, when whole network is divided into multiple clusters each cluster has one CH. This CH is responsible for communication of all nodes in the cluster. CH receives data from all nodes at single-hop, aggregates and transmits directly to BS or through intermediate CHs. In Multi-hop inter-cluster communication when distance between CH and BS is large then CH use intermediate CHs to communicate to BS.

Fig. 2 describes Multi-Hop LEACH communication architecture. Randomized rotation of CH is similar to LEACH. Multi-Hop LEACH selects best path with minimum energy consuming route. An other criteria of selecting intermediate CH is to keep overall distance towards BS should be minimum because distance is directly proportional to energy dissipation. So, a route with minimum hop-count between source CH and BS is selected.

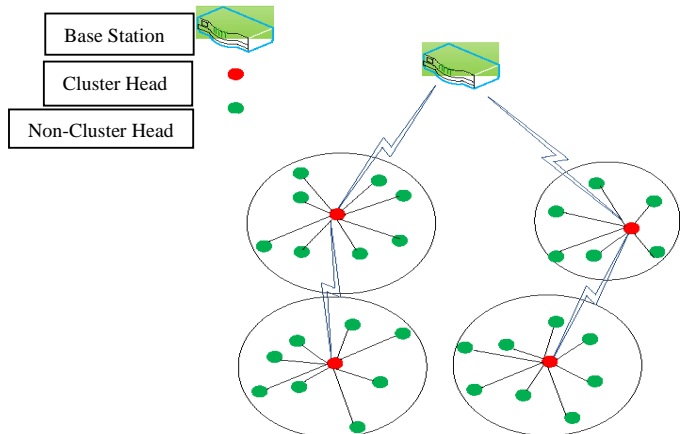

Fig. 2 Multi-Hop LEACH

### II.3.4    M-LEACH

LEACH considers all nodes as fixed and homogeneous with respect to their energy, which is not a realistic approach. In particular round uneven nodes are attached to multiple CH. In this case CHs with large number of member nodes will drain their energy very quickly as compare to CHs with smaller number of member nodes associated. Mobility support is also very important issue faced by LEACH routing protocol. To mitigate these issues, M-LEACH has been proposed in [16].

M-LEACH allows mobility for all nodes during the setup and steady state phase. Some assumptions are also made in M-LEACH like other routing protocols. Initially all nodes are homogeneous in sense of antenna gain, all nodes have their location information through Global Positioning System (GPS) and BS is considered fixed in M-LEACH. Distributed setup phase of LEACH is modified by M-LEACH in order to select most suitable CHs. M-LEACH also considers remaining energy of the nodes in selection of CHs. In M-LEACH CHs are elected on the basis of attenuation model [17]. Other criteria for CH selection is speed of mobility. Nodes with minimum mobility and lowest attenuation power are selected as CHs. Then selected CHs broadcast their status to all nodes in their transmission range. Nodes compute their willingness from multiple CHs and select the CH with maximum residual energy.

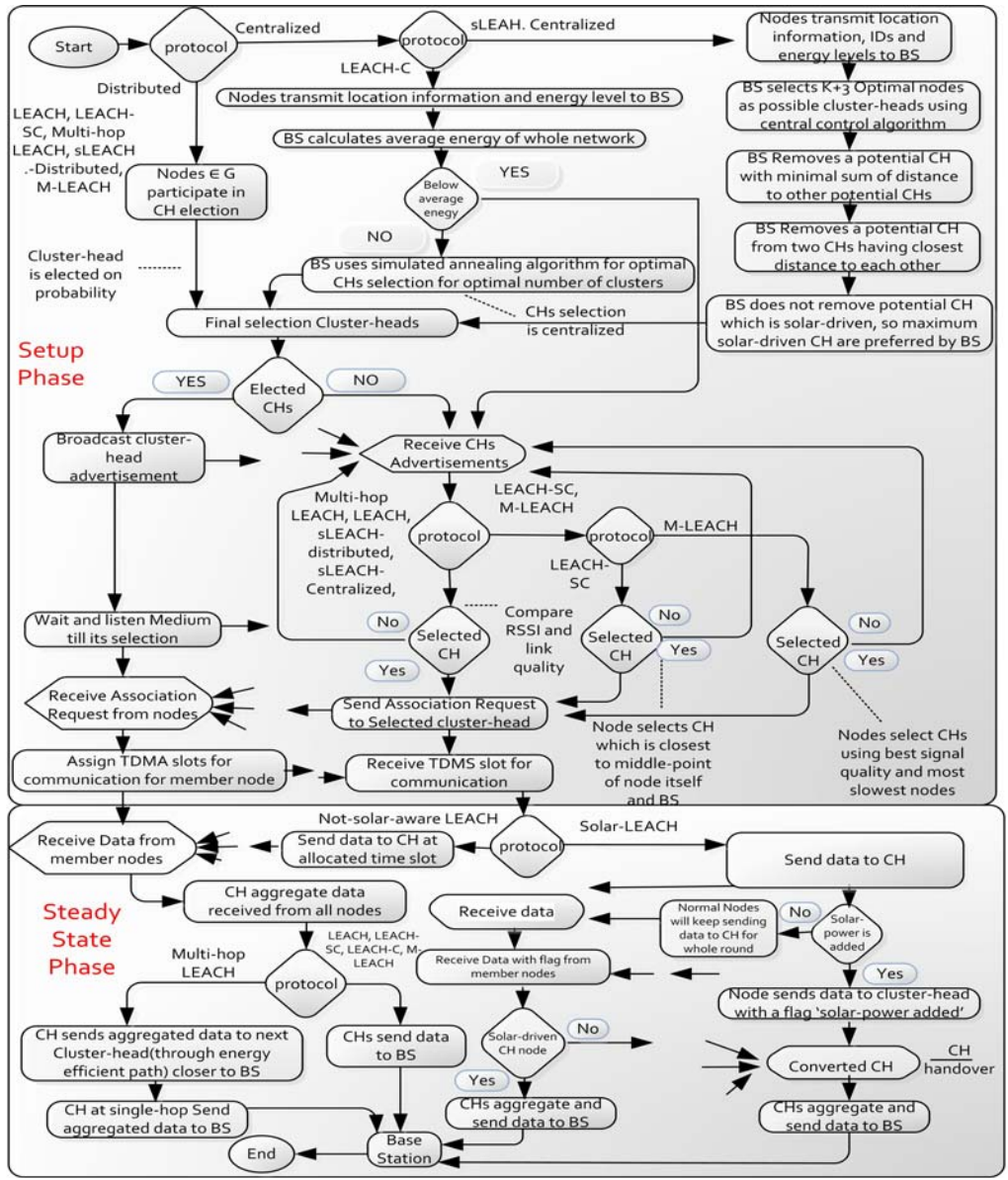

**Fig. 3** A combined flow chart of all routing protocol

In steady state phase, if nodes move away from CH or CH moves away from its member nodes than any other CH will become suitable for member nodes. This phenomena results into inefficient clustering formation. To deal with this problem M-LEACH provides handover mechanism for nodes to switch on to new CH. When nodes decide to make handoff, then nodes send DIS-JOIN message to current CH and also send JOIN-REQ to new CH. After handoff, CHs re-schedule the transmission pattern for member nodes.

### II.3.5      LEACH-SC

In earlier clustering routing protocols, authors address the position of nodes with respect to their CH and BS to some extent. But LEACH-SC proposed in [8], deals with this phenomena and provides the reasonable solutions about their relative distance and position. Actually energy dissipation depends upon the relative position and distance among non-CHs, CHs and BS. Clustering protocols basically try to minimize the distance of transmission among normal nodes to CHs and CHs to BS. But some time nodes are sending data to their CH in opposite direction of BS. In this scenario data is transmitted with additional distance. LEACH-SC addresses these kinds of issues in order to save the transmitting energy cost of the sensor nodes and improves the network's lifetime .

Operation of LEACH-SC is based on rounds. Each round is consisting of setup phase and steady state phase. But LEACH-SC slightly alters the clustering formation. In improved clustering formation algorithm of LEACH- SC, selected CHs advertise their IDs and location information to all nodes in range. Nodes receive these information from all CHs within communication range. Nodes compare information and select their CHs which is nearest to the middle-point between BS and comparing non-CH node itself. Basically in this improved clustering formation algorithm, authors change the way of making membership between non-CHs nodes and selected CHs. A combined flow chart of all is shown in Fig 3.

### III. CLASSIFICATION AND COMPARISON OF LEACH AND ITS MODIFIED ROUTING PROTOCOLS IN WIRELESS SENSOR NETWORKS

All routing protocols have some significant properties and address specific issues to produce some betterment in existing routing protocols. Each routing protocol has some advantages and capabilities. Routing protocols face some common energy dissipation challenges e.g., Cost of Clustering, Selection of CHs and Clusters, Synchronization, Data Aggregation, Repair Mechanisms, Scalability, Mobility, and initial energy of all nodes [14]. We compare above mentioned routing protocols with respect to some very important performance characteristics for WSNs. These characteristics of WSNs are following.

**Table 1** Performance comparison of hierarchical routing protocol

| Clustering Routing protocol | Classification | Mobility | Scalability | Self-organization | Randomized rotation | Distributed | Centralized | Hop-count | Energy efficiency | Resources awareness | data aggregation | homogeneous |
|---|---|---|---|---|---|---|---|---|---|---|---|---|
| LEACH | Hierarchical | fixed BS | limited | Yes | Yes | Yes | No | Single-hop | High | Good | Yes | Yes |
| (LEACH-Centrlized) | Hierarchical | fixed BS | Good | Yes | yes | No | Yes | Single-hop | High | Good | Yes | Yes |
| (sLEACH-Centralized) | Hierarchical | fixed BS | Good | Yes | Yes | No | Yes | Single-hop | Very High | Very Good | Yes | Yes |
| (sLEACH-Distributed) | Hierarchical | fixed BS | Good | Yes | Yes | Yes | No | Single-hop | Very High | Very Good | Yes | Yes |
| (Multi-Hop LEACH) | Hierarchical | fixed BS | Very Good | Yes | Yes | Yes | No | mult-hop | Very High | Very Good | Yes | Yes |
| (M-LEACH) | Hierarchical | Mobile BS and nodes | Very Good | Yes | Yes | Yes | No | sinle-hop | Very High | Very Good | Yes | Yes |
| LEACH-SC | Hierarchical | fixed BS | Good | Yes | Yes | Yes | No | Single-hop | High | Good | Yes | Yes |

- Classification: The classification of a routing protocol indicates that it is flat, location-based or hierarchal routing protocol [15].
- Mobility: It specifies that routing protocol support mobility or not.
- Scalability: It describes how much routing protocol is scalable, if the network density is increased.
- Randomized Rotation of CHs: Randomized Rotation of CH is very necessary in order to drain the battery of all nodes equally [1].
- Distributed clustering algorithm: CHs are self-elected in distributed clustering algorithm [1].
- Centralized clustering algorithm: CHs are selected by BS, using central control algorithm [3].
- Single-hop or Multi-hop: It is also important feature of routing protocol.

- Single-hop is energy efficient if it is smaller area of network and multi-hop is better for denser network [11].
- Energy Efficiency: It is the main concern of energy efficient routing protocol to maximize lifetime of the network [1], [2], [4], [11], [15].
- Data Aggregation: In order to reduce amount of data transmitted to BS, CHs perform data-aggregation in this way CHs transmission energy cost is reduced [1], [2].
- Homogeneous: Homogeneity of all nodes is considered in some routing protocols which describes that initial energy level of all the nodes is same.

Table.I shows the comparison LEACH, LEACH-C, sLEACH, M-LEACH, Multi- Hop LEACH and LEACH-SC. Performance comparison shows that behavior of these routing protocols is similar in many ways. All routing protocol are hierarchal, homogeneous, perform Data aggregation, self-organization, ran- domized rotation of CHs and having fixed BS despite M-LEACH. LEACH, LEACH-SC, M-LEACH and Multi-Hop LEACH use distributed algorithm for CH selection. LEACH-C uses central control algorithm for CH selection and sLEACH is designed for both centralized and distributed algorithm. LEACH, sLEACH, LEACH-SC and M-LEACH are routing protocol in which BS is at single-hop and in Multi-Hop LEACH BS can be at multi-hop distance from CHs. LEACH and M-LEACH allow limited scalability. LEACH-C, sLEACH and LEACH-SC allow good scalability while, Multi-Hop LEACH is providing maximum scalability feature due to multi-hop communication option for CHs.

## IV. ANALYTICAL COMPARISON FOR ENERGY EfficIENCY OF ROUTING PROTOCOLS

For analytical comparison, it is essential to be aware of radio model character- istics adopted by energy efficient routing protocols. All energy efficient routing protocols proposed in previous research provide different characteristics about the radio distinctiveness. These different characteristics cause significant varia- tion in energy efficiency of routing protocols. Radio model differentiates energy dissipation to run transmitter and receiver circuitry per bit. Radio transceiver dissipates $\varepsilon$ amp for transmission amplifier to attain suitable $E_b/N_0$ [1].

Table 2 Radio Transceiver Characteristics

| Operation | Energy Dissipation |
|---|---|
| Transmitter Electronics (EelectTx) | 50 nj/bit |
| Receiver Electronics (EelecRx) | 50 nj/bit |
| Transmit amplifier (Eafs) | 100 pj/bit/m2 |

Multiple radio model's standard have been declared for different type of sensor nodes. Radio characteristics of radio model, which are used in most of the sensor's literature is shown in table 2. Transmitter and receiver Radio model is shown in Fig. 4. Mainly energy dissipation of a individual node depends upon number of transmissions, num- ber of receiving, amount of data to transmit, distance between transmitter and receiver. So, first we describe the possible source of energy consumption. Then we will compare selected routing protocols. We will also analyze how energy efficiency is achieved by these routing protocols.

### IV.1 ENERGY CONSUMPTION

There are multiple source of energy consumption and every energy efficient routing protocol deals in different manner to reduce energy consumption. This section provides mathematical analysis of possible energy consumption sources. In most of the cases in WSNs, free space communication model is assumed among all nodes and BS.

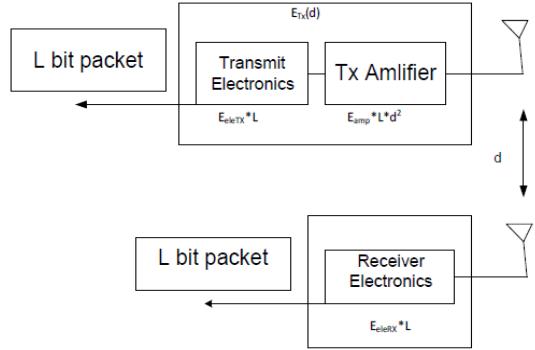

Fig. 5 Linear Network Model

As we know that in cluster-based routing protocols, duplex communication is needed, especially for some query-based applications. During upward com- munication nodes send their data to CH and CH forwards aggregated data to BS. So energy consumption on CH will be:

$$E_{CHup}=(n/K-1)(E_{eleRX}\times L_C)+n/K\times L_C\, E_{AD}+E_{eleTX}\times L_A+E_{amp}\times L_A\times d_{toBS}2 \quad (3)$$

where, $n$ is number of all nodes in WSNs, $ECHup$ is total upward communication energy consumption of Cluster-head, $((n/k)-1)$ number of possible nodes in one cluster, $K$ is possible number of clusters, $LC$ is data of non-CH nodes, $EAD$ is energy cost for data aggregation, $LA$ is aggregated data, $d^2$ to BS is distance between CH and BS.

Energy consumption of single non-CH node will be:

$$E_{nCHup} = E_{eleTX}\times L_C + E_{afs}\times L_C \times d^2_{toCH} \quad (4)$$

Where $d^2_{toCH}$ is distance between CH and member node. Energy consumption of all the nodes in one cluster will be:

$$E_{nCHup} \times ((n/k)-1) = ((n/k)-1)(E_{eleTX}\times L_C + E_{afs}\times L_C \times d_{toCH}) \quad (5)$$

So total upward energy cost of single cluster will:

$$E_{up}= E_{CHup} + E_{nCHup} \times ((n/k)-1) \quad (6)$$

When BS has to get specific sensing information from nodes, in this case BS sends instructions to CHs only. CHs send these instructions to member nodes. In this process CHs and non-CHs nodes also pay energy cost. This downward energy cost is not considered in some cases, but it is certainly paid. Energy consumption is not a problem for BS so, energy consumption of BS is ignored. If BS sends instructions for all nodes, then energy consumption on CH will be:

$$E_{CHdown}=(n/K)E_{eleRX}\times L_{BS}+(n/K-1)x(E_{eleTX}+E_{amp}\times L_{BS}\times d_{tonCH}2) \quad (7)$$

When CH transmits to its member nodes, receiving nodes also consume energy and it will be equal to:

$$E_{nCHdown}((n/k)-1) = ((n/k)-1)(E_{eleRX}\times L_{BS}) \quad (8)$$

Total downward energy consumption will be:

$$E_{down}= E_{nCH} + E_{CH} \quad (9)$$

So total estimated energy consumption for duplex communication of a single cluster will be:

$$E_C = E_{up} + E_{down} \quad (10)$$

From equation 10 total energy consumption of whole network can be also be estimated, and it will be:

$$E_T = E_C \times K \quad (11)$$

Clustering routing protocols for WSNs also bear energy cost in setup phase. Every node keep sensing continuously which also cause energy dissipation . These kind of energy dissipations are not considered mostly. Only transmission energy dissipation is compared in analytical comparison.

**IV.1   ENERGY EFFICIENCY OF CLUSTERING ROUTING PROTOCOLS**

We compare the selected Hierarchical routing protocols in order to analyze energy efficiency. We only consider upward transmission energy dissipation of each routing protocol for comparison. In this scenario all nodes have to trans- mit their data to BS through multiple CHs. Distance between nodes and CHs can plays key rule in amount of energy dissipation. LEACH reduces energy dissipation over a factor of 7x and 8x reduction as compared to direct commu- nication and a factor of 4x and 8x compared to the Minimum Transmission Energy (MTE) [1]. This energy efficiency is due to reduction of number of direct transmissions because, in LEACH only CHs directly communicate to BS and remaining nodes have to transmit to CHs which is at smaller distance.

The LEACH-Centralized is improved form of LEACH and enhance the network lifetime. Optimal number of CHs and distance reduction between CH and normal nodes is the main factor in order to increase network lifetime. The sLEACH provides better network lifetime as compare to LEACH and other compared routing protocols. It is because of CH selection is not uniform in sLEACH. In sLEACH solar-aware nodes have more probability to be selected as CHs as compare to battery-driven nodes. Multi-Hop LEACH is more energy efficient than LEACH [11].

Multi-Hop LEACH also provides better connectivity and more successful data transmission as compare to LEACH [12]. The reason behind this en- hancement is multi-hop communication adopted by CHs. As member nodes save energy by sending data to CH instead of BS. Similarly in Multi-Hop LEACH CH at longer distance from BS, transmits data to next CH closer to BS instead of direct transmission to BS. Multi-Hop LEACH is more effective when network diameter is larger. Energy efficiency of multi-hop-LEACH can be better elaborated with the example of linear network shown in Fig 5. In this network, two CHs A and B are communicating to BS. A uniform distance 'm' is taken among BS and two CHs.

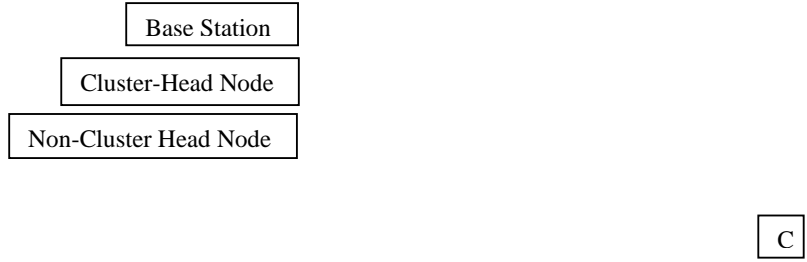

Fig. 5 Linear Network Model

In order to calculate the transmitting energy cost of CHs A and B, which are directly transmitting to BS will be:

$$E_{LEACH} = E_{eleTX} \times L_A + \epsilon_{afs} \times L_A \times 2m$$
$$+ E_{eleTX} \times L_B + \epsilon_{afs} \times L_B \times m^2 \quad (12)$$

where, $E_{LEACH}$ is total energy cost of CH A and B, $L_A$ is aggregated data transmitted by CH A and $L_B$ is aggregated data transmitted by CH B towards BS and $m$ is equal distance among CHs and BS.

Similarly total transmitting energy cost can also be calculated when multi- hop communication is taking place. Multi-hop LEACH utilizes multi-hop com- munication. In this linear network, if CH A transmits data to CH B instead of BS, then CH B has to transmit not only its own cluster's data but also has to transmit CH B's data to BS.

$$E_{mhL} = E_{eleTX} \times L_A + \epsilon_{afs} \times L_A \times m^2 + E_{eleRX}$$
$$\times L_A + E_{eleTX} \times (L_B + L_A) + \epsilon_{afs} \times (L_B + L_A) \times m^2 \quad (13)$$

Where, $E_{mhL}$ is total transmitting energy cost of both CHs in case of multi-hop communocation of Multi-hop LEACH. Cluster-head near BS has more traffic burden in case of M-LEACH. But CH which is at longer distance from BS has benefits because it has to transmit at small distance and increase its lifetime. As equation 12 and 13 are measuring that Multi-hop

LEACH is consuming more energy apparently, that's why it is very important to know that Multi-hop LEACH will be only effective when the network is very large and some CHs are at longer distance from BS. Otherwise LEACH and other routing protocols allowing single-hop communication between CH and BS will be more effective. LEACH-SC is also more energy efficient as compare to LEACH. As we know distance is directly proportional to energy consumption and LEACH-SC is more efficient because it minimize the backward communication of all nodes with respect to BS.

## V. SIMULATION RESULTS AND ANALYSIS

In this section we present simulation results of LEACH, LEACH-C, Multi-hop LEACH, M-LEACH, sLEACH-Centralized, sLEACH-Distributed and LEACHSC to make effective and critical analysis. Simulation parameters are shown in Table 3. This simulation is done by using MATLAB. 100 nodes are scattered uniformly in region of 100*m*_100*m*. During simulation of these routing protocols we adjusted the network topology according to realtime behavior of sensors nodes and also consider re-energization ability of solar-driven sensors in sLECAH nodes to obtain more realistic simulation results.

| Parameter | value |
|---|---|
| Network size | 100m * 100m |
| Initial Energy | .5 j |
| p | .1 j |
| Data Aggregation Energy cost | 50pj/bit j |
| Number of nodes | 100 |
| Packet size | 200 bit |
| Transmitter Electronics (EelectTx) | 50 nj/bit |
| Receiver Electronics (EelecRx) | 50 nj/bit |
| Transmit amplifier (Eamp) | 100 pj/bit/m2 |

Table 3 Simulation Environment

Fig.6 shows the network lifetime. In LEACH all nodes reach to death firs, then LEACH-C, M LEACH, LEACH-SC, Multi-hop LEACH, sLEACHCentralized and then sLEACH-distributed respectively. In solar-aware LEACH routing protocols nodes die after longest period of time because solar-awre nodes have ability to re-energize themselve for certain period. The sLEACH has 300 % more network lifetime as compare to LEACH, because in sLEACH last node is dying after 4000 rounds. This sLEACH efficiency can also be improved by adding more solar-driven nodes as compare to battery driven nodes.

The sLEACH-Distributed is slightly better then sLEACH-Centralized because in sLEACH-Distributed localized clustering formation is performed. LEACH-C has almost 23 % better network lifetime as compare to LEACH and LEACH-SC has almost 33 % better network lifetime as compare to LEACH. M-LEACH has 30 % better network lifetime as compare to LEACH because last node reaches to death after 500 rounds. In Multi-hop LEACH routing protocol produces almost 90 % network life enhancement as compare to LEACH and it can be further improved if the network diameter is increases.

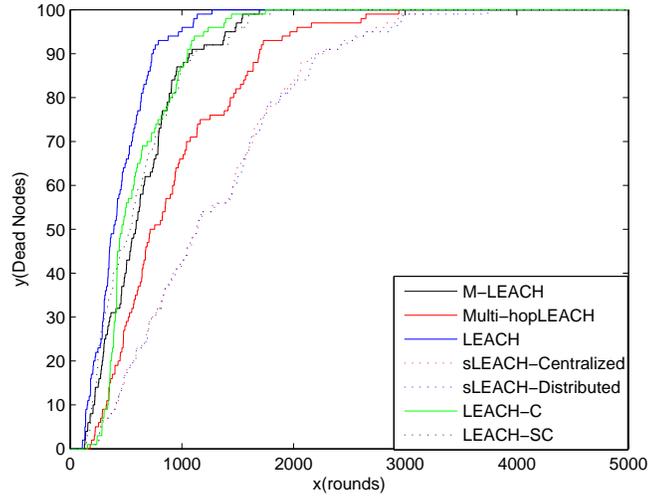

Fig. 6  Number of dead nodes

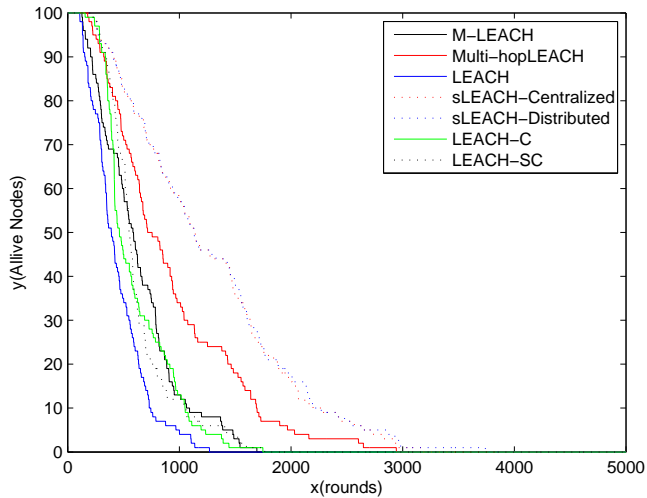

Fig. 7  Number of alive nodes

Fig. 7 shows the number of allive nodes with respect to number of rounds for all selected routing protocols. Till 500 rounds all nodes are alive for every routing protocol. Stable period in which all nodes are alive, is maximum in LEACH-C and sLEACH-Centralized as compare to all other distributed routing protocols. LEACH has 28 %, 33 %, 35 %,120 % and 300 % less survival time as compare to M-LEACH, Multi-hop LEACH, and sLEACH respectively. Reason is similar as describe for Fig 6. Data ( actual data signal) successfully transfer to BS indicates the quality of a routing protocol. If data received by BS is increasing it means quality of routing protocol is getting better and better. Fig 8 shows the quality analyzing of clustering routing protocols. As CHs are responsible for aggregating and transmitting data to BS, so routing protocol with optimum number of CHs will be more efficient. Multi-hop LEACH has better quality than LEACH because, CHs at the corner of the network have to transmit to next CHs towards BS while in LEACH CHs have to transmit directly to BS at longer distance. It will result into poor signal strength and less successful data delivery.

M-LEACH also provides better quality in dynamic topology of network. Comparatively sLEACH has maximum quality because, in sLEACH CHs are elected on the basis of solar property of nodes. Maximum CHs in sLEACH are solar-driven nodes and these CHs serve for longer period of time as compared to battery-driven CHs in sLEACH. These solar-driven CHs have enough energy to transmit at longer distance with acceptable signal strength, that's why sLEACH has maximum quality of network. In sLEACH, sLEACH-Distributed has more quality as compare to sLEACH-Centralized as shown in Fig. 8. It is because of increasing probability of solar-driven nodes to be CHs and its proved by equation 2 of this paper. LEACH has minimum data transmitted to BS as compare to all other routing protocols. LEACH-C has significant performance because optimal and guaranteed CHs are selected for every round. LEACH-SC, M-LEACH ans Multi-hop LEACH is also performing better as compare to LEACH**.**

As data to BS is important factor for quality analysis of any routing protocol, similarly data(data signal) to CH is also important. Fig 9 shows the data received by CH. Results are similar as we computed from Fig 8. however, LEACH-C is not as efficient in data to CHs as in Data to BS. It is

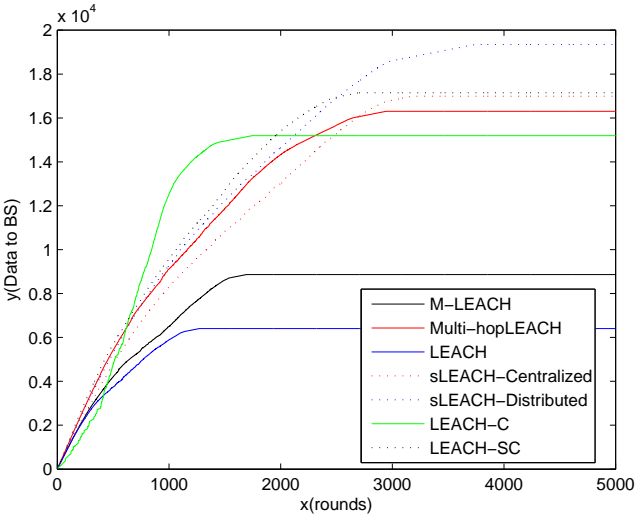

Fig. 8 Number of data packets received at the BS

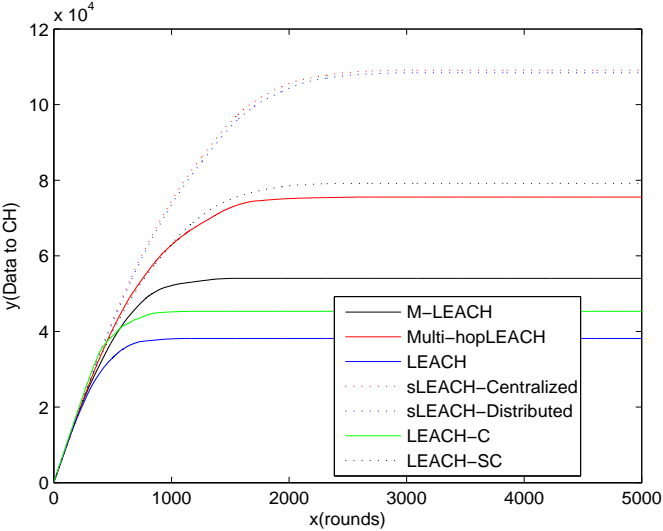

Fig. 9 Number of data packets received at the Cluster-head

because of normal nodes are reduced due to consistent percentage of CHs for all rounds. In this case sLEACH-Centralized is also very much better than LEACH. Other improved routing protocols M-LEACH, Multi-hop LEACH and LEACH-SC also deliver more data to CHs. As

LEACH, M-LEACH, LEACH-SC, Multi-hop LEACH and sLEACHDistributed use distributed self-organization algorithm, because of this optimal mal number of CHs are not guaranteed. Fig. 10 shows uncertain number of CHs elected per rounds. Results shows that LEACH, Multi-hop LEACH, MLEACH, LEACH-SC and sLEACH-Distruted show more uncertainty as compare to other routing protocols. sLEACH-Distributed is slightly better incase of CHs selection because criteria of randomized rotation of CHs is modified from LEACH. LEACH-C and sLEACH-Centralized use central control algorithm that's why number of CHs are specific and provide efficient clustering creation. Uncertainty about CHs selection for Distributed hierarchical routing protocols is decreasing the performance of these protocols.

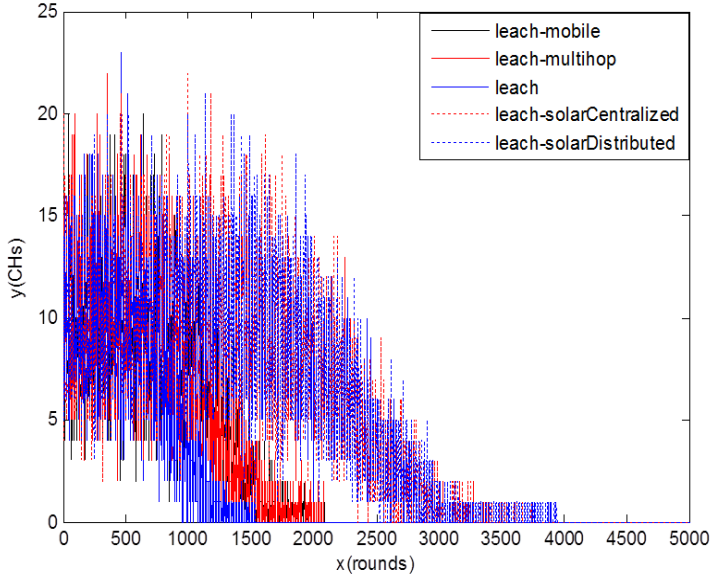

Fig. 10   CHs per round

## VI.    CONCLUSION AND FUTURE WORK

In this paper we discussed LEACH, LEACH-C, Multi-hop LEACH, M-LEACH Solar-aware LEACH (sLEACH) and LEACH-SC hierarchical routing protocols for WSNs. The main concern of this research is to examine the energy management efficiency and throughput enhancement of these routing protocols. We compare the lifetime and data delivery characteristics with the help of analytical comparison and also from our simulation results. As simulation results indicate that energy harvesting technique for example sLEACH is providing maximum energy efficiency and quality of service. In future work these energy harvesting techniques should be the area of interest. Significant research work has been done in these different clustering routing protocols in order to increase the lifetime and data delivery features. Certainly further energy improvement is possible in future. Distributed and centralized algorithms should be developed to enhance the setup phase all routing protocols. Improvement is also possible in many aspects like sensor nodes electronics, nodes deployment management, effective and energy efficient routing protocols selection for WSNs according to requirements of application.

In future, we are interested to deal with energy efficiency at MAC layer like [33-34-35-36], the effects of antenna orientation with respect to human body like [37-38], and applicability of these protocols under heterogeous environment ([39-40-41-42]) are also under consideration.